\documentclass[%
aps,
prd,
reprint,
showpacs
]{revtex4-1}

\usepackage[dvipsnames]{xcolor}

\usepackage{mathtools} 
\usepackage{amsthm}
\usepackage{amsfonts}
\usepackage{amssymb}
\allowdisplaybreaks

\usepackage{tensor}
\newcommand{\ind}{\indices}

\usepackage[T1]{fontenc}
\usepackage{lmodern}

\usepackage[english]{babel}
\def\Ad{\mathrm{Ad}}

\begin{document}

\title{Cartan geometry of spacetimes with a nonconstant 
	cosmological function $\Lambda$}
\author{Hendrik Jennen}
\email{hjennen@ift.unesp.br}
\affiliation{%
	Instituto de F\'{\i}sica Te\'orica, UNESP-Universidade 
	Estadual Paulista, \\
	Rua Dr.~Bento Teobaldo Ferraz, 271 - Bl.~II, 01140-070, S\~ao 
	Paulo, SP,  Brazil
}
\date{\today}

\begin{abstract}
	We present the geometry of spacetimes that are tangentially 
	approximated by de~Sitter spaces whose cosmological constants 
	vary over spacetime. Cartan geometry provides one with the 
	tools to describe manifolds that reduce to a homogeneous Klein 
	space at the infinitesimal level. We consider a Cartan 
	geometry in which the underlying Klein space is at each point 
	a de~Sitter space, for which the combined set of pseudo-radii 
	forms a nonconstant function on spacetime.  We show that the 
	torsion of such a geometry receives a contribution that is not 
	present for a cosmological constant. The structure group of 
	the obtained de~Sitter-Cartan geometry is by construction the 
	Lorentz group~$SO(1,3)$.  Invoking the theory of nonlinear 
	realizations, we extend the class of symmetries to the 
	enclosing de~Sitter group~$SO(1,4)$, and compute the 
	corresponding spin connection, vierbein, curvature, and 
	torsion.
\end{abstract}

\pacs{04.20.Cv, 02.40.-k, 98.80.-k}

\maketitle

\section{Introduction}
\label{sec:intro}

In theories of gravity, the strong equivalence principle implies 
that spacetime $\mathcal{M}$ is locally approximated by the 
spacetime underlying special relativity~\cite{DiCasola:2013nep}.
Since the laws that govern special relativistic physics are 
covariant with respect to the Poincar\'e group $ISO(1,3)$, the 
corresponding spacetime is the affine Minkowski space~$M$.
Although finite Poincar\'e translations are not defined for a 
generic spacetime, the equivalence principle indicates that 
locally they are in one-to-one correspondence with infinitesimal 
active diffeomorphisms, for both sets generate translations along 
spacetime~\cite{Hehl:1976grs}. Mathematically speaking, there is 
a $1$-form, called the vierbein, which at any point
is valued in the algebra of Poincar\'e translations $\mathfrak{t} 
= \mathfrak{iso}(1,3)/\mathfrak{so}(1,3)$. The vierbein pulls 
back or \emph{solders} the geometric and algebraic structure of 
$\mathfrak{t}$ to spacetime.  For example, the Minkowski metric 
on $\mathfrak{t}$ gives way to a metric of the same signature on 
$\mathcal{M}$, from which it follows that the vierbein can be 
chosen to be an orthonormal frame---an idealized observer---along 
spacetime.  Due to the equivalence principle, Lorentz 
transformations of these observers constitute a symmetry and are 
therefore elements of the structure group of the geometry, which 
in turn leads to the introduction of a spin connection.

The right mathematical framework for the setting just outlined is 
due to Elie Cartan~\cite{Cartan:1926gh}, in which the 
$\mathfrak{so}(1,3)$-valued spin connection and the
$\mathfrak{t}$-valued vierbein are combined into an
$\mathfrak{iso}(1,3)$-valued Cartan connection, thereby defining 
a Riemann-Cartan geometry~\cite{Hehl:1976grs}. It is explained 
comprehensibly in~\cite{Wise:2010sm} how the 
$\mathfrak{iso}(1,3)$-valued connection gives a prescription for 
\emph{rolling without slipping} the affine Minkowski space along 
the integral curves of vector fields on spacetime.
It is indeed the central idea behind Cartan geometry that a 
homogeneous model space is generalized to a nonhomogeneous space, 
for which the local structure is algebraically isomorphic to the 
one of the model space~\cite{sharpe1997diff_geo}, and where the 
degree of nonhomogeneity is quantified by the presence of 
curvature and torsion.  In the manner thus explained, the choice 
for a Riemann-Cartan geometry to describe spacetimes underlying 
theories of gravity is implied by the equivalence principle, 
together with the assumption that the local kinematics are 
governed by the Poincar\'e group.

When the $\mathfrak{iso}(1,3)$-valued Cartan connection is 
replaced by one that is valued in the de~Sitter algebra 
$\mathfrak{so}(1,4)$, spacetime is locally approximated by a 
de~Sitter space $dS$ in place of the affine Minkowski space, a 
structure we shall call a de~Sitter-Cartan geometry.  Since the 
vierbein is valued in the space of de~Sitter transvections 
$\mathfrak{p} = \mathfrak{so}(1,4)/\mathfrak{so}(1,3)$, 
translations in a de~Sitter-Cartan spacetime are generated by 
elements of $\mathfrak{p}$. This implies that the commutator of 
infinitesimal translations is proportional to a Lorentz rotation.  
The constant of proportionality is essentially the cosmological 
constant of the tangent de~Sitter spaces~\cite{Wise:2010sm}. It 
is then sensible to identify this geometric cosmological constant 
with the dark energy on spacetime. Such an interpretation is in 
concordance with the MacDowell-Mansouri model for 
gravity~\cite{MacDowell1977}. In this model, the fundamental 
field  is indeed a $\mathfrak{so}(1,4)$-valued Cartan connection, 
for which the action is equivalent, up to topological terms, with 
the Palatini action for general relativity in the presence of a 
cosmological constant~\cite{Wise:2010sm,Westman:2012xk}.

At any point in a de~Sitter-Cartan spacetime, the cosmological 
constant is related to a length scale defined in the commutation 
relations of the de~Sitter transvections. Therefore, it is rather 
straightforward to generalize to geometries in which this length 
scale becomes a nonconstant function on spacetime. In 
Sec.~\ref{sec:lin_dSC_geo}, we claim some originality for 
constructing a de~Sitter-Cartan geometry that provides spacetime 
with a \emph{cosmological function} $\Lambda$, which in general 
does not satisfy $d\Lambda = 0$. We shall see that a nonconstant 
$\Lambda$ gives rise to a new term in the expression for the 
torsion of the de~Sitter-Cartan geometry.

\section{de~Sitter-Cartan geometry with a cosmological function}
\label{sec:lin_dSC_geo}

A de~Sitter-Cartan geometry is the Cartan geometry modeled on 
$(\mathfrak{so}(1,4),SO(1,3))$, which means it consists of a 
principal Lorentz bundle $P(\mathcal{M},SO(1,3))$ over spacetime, 
on which is defined a $\mathfrak{so}(1,4)$-valued Cartan 
connection $A$.  For a rigorous discussion on Cartan geometry, 
see~\cite{sharpe1997diff_geo,Alekseevsky:1995cc}, while the 
articles~\cite{Wise:2010sm,Wise:2009fu,Westman:2012xk} are very 
helpful to develop an intuition that goes along with the 
mathematics. The connection $A$ provides spacetime with the 
information that it is tangentially approximated by de~Sitter 
space, the homogeneous Klein space with respect to which 
inhomogeneities are measured~\cite{Klein:1893}. We shall 
construct a de~Sitter-Cartan geometry, in which these tangent 
de~Sitter spaces have cosmological constants that are not 
required to be the same over spacetime.  As a consequence, the 
thus obtained geometry describes a manifold with arbitrary 
curvature and torsion on which a nonconstant cosmological 
function $\Lambda$ is defined from the onset.

Under the action of a local $SO(1,3)$-transformation $h$, the 
de~Sitter-Cartan connection transforms according 
to~\cite{sharpe1997diff_geo}
\begin{equation}
	A \mapsto \Ad(h) ( A + d ).
\end{equation}
The connection is valued in the de~Sitter algebra 
$\mathfrak{so}(1,4)$, which is characterized by the commutation 
relations
\begin{gather}
\label{eq:comm_relations_so(1,4)}
\begin{aligned}
	-i[M_{ab},M_{cd}] &= \eta_{ac}M_{bd} - \eta_{ad}M_{bc} + 
	\eta_{bd}M_{ac} - \eta_{bc}M_{ad}, \\
	-i[M_{ab},P_c] &= \eta_{ac}P_b- \eta_{bc}P_a, \\
	-i[P_a,P_b] &= -l^{-2}M_{ab},
  \end{aligned}
\end{gather}
where $\eta_{ab} = (+,-,-,-)$, while we parametrize an element of 
$\mathfrak{so}(1,4)$ by $\tfrac{i}{2} \lambda^{ab} M_{ab} + i 
\lambda^a P_a$. The reductive nature of the algebra schematically 
reads as
\begin{equation}
	\label{eq:red_split_so(1,4)}
	\mathfrak{so}(1,4) = \mathfrak{so}(1,3) \oplus \mathfrak{p},
\end{equation}
where $\mathfrak{so}(1,3) = \mathrm{span}\{M_{ab}\}$ is the 
Lorentz subalgebra and $\mathfrak{p} = \mathrm{span}\{P_a\}$ the 
subspace of infinitesimal de~Sitter transvections or 
translations. The latter are defined by $P_a = M_{a4}/l$, where 
$l$ is an \emph{a-priori} arbitrary length scale that effectively 
determines the cosmological constant of the corresponding Klein 
geometry $dS = SO(1,4)/SO(1,3)$, namely,~\cite{Wise:2010sm}
\begin{equation}
	\label{eq:rel.Ll}
	\Lambda = \frac{3}{l^2}.
\end{equation}

Since the Cartan connection is at any point $x \in \mathcal{M}$ 
valued in a copy of $\mathfrak{so}(1,4)$, we may choose the set 
of length scales $l(x)$ to form a smooth function.  Doing so, the 
cosmological constants of the corresponding tangent de~Sitter 
spaces also constitute a generic nonconstant cosmological 
function $\Lambda(x)$ on spacetime. In the following paragraphs 
we discuss the implications of a nonconstant $\Lambda$ for the 
de~Sitter-Cartan geometry.

Corresponding to the reductive splitting 
\eqref{eq:red_split_so(1,4)}, we decompose the Cartan connection 
and its curvature $F = dA + \tfrac{1}{2}[A,A]$ as
\begin{equation}
	\label{eq:red_split_AF_dSC}
	A = \tfrac{i}{2} A^{ab} M_{ab} + iA^a P_a
	~\text{and}~
	F = \tfrac{i}{2} F^{ab} M_{ab} + iF^a P_a.
\end{equation}
from which it follows that $A^a$ and $F^a$ have the dimension of 
length.  The $\mathfrak{so}(1,3)$-valued $1$-form $A^{ab}$ is an 
Ehresmann connection for local Lorentz 
transformations~\cite{Alekseevsky:1995cc}, i.e.,~a spin 
connection, while the forms $A^a$ constitute a vierbein. Note 
that the decompositions~\eqref{eq:red_split_AF_dSC} are well 
defined, since local Lorentz transformations leave the reductive 
splitting invariant.  Due to the presence of a spin connection 
and vierbein, it is possible to define local Lorentz and 
diffeomorphism covariant differentiation, as well as a metric 
structure on spacetime; see, e.g.,~\cite{Ortin:2004}.

Given the commutation
relations~\eqref{eq:comm_relations_so(1,4)}, one computes the 
curvature~$F^{ab}$ and torsion~$F^{a}$ in terms of the spin 
connection and vierbein:
\begin{subequations}
	\label{eqs:curvtors_dSC}
\begin{align}
	\label{eq:curv_dSC}
	F^{ab} &= dA^{ab} + A\ind{^a_c} \wedge A^{cb} + \frac{1}{l^2} 
	A^a \wedge A^b \\
	\nonumber
	&= d_A A^{ab} + \frac{1}{l^2} A^a \wedge A^b,
	\\
	\label{eq:tors_dSC}
	F^a &= dA^a + A\ind{^a_b} \wedge A^b - \frac{1}{l} dl \wedge 
	A^a 
	\\
	\nonumber
	&= d_A A^a - \frac{1}{l} dl \wedge A^a.
\end{align}
\end{subequations}
In the limit of an everywhere diverging length scale $l$, or 
equivalently, an everywhere vanishing cosmological constant, the 
expressions~\eqref{eqs:curvtors_dSC}
reduce to the curvature~$d_A A^{ab}$ and torsion~$d_A A^a$ for a
Riemann-Cartan geometry~\cite{Ortin:2004}. In the generic case, 
however, the curvature and torsion are not given by the exterior 
covariant derivatives of the spin connection and vierbein. The 
extra term in~\eqref{eq:curv_dSC} represents the curvature of the 
local de Sitter space. This contribution is present because the 
commutator of two infinitesimal de~Sitter transvections equals an 
element of the Lorentz algebra. In addition, there is a new term 
in the expression~\eqref{eq:tors_dSC} for the torsion if the 
length scale is a nonconstant function.  This term comes about as 
follows. The torsion is the $\mathfrak{p}$-valued $2$-form 
$F_\mathfrak{p} = dA_\mathfrak{p} + 
[A_{\mathfrak{h}},A_\mathfrak{p}]$~\cite{sharpe1997diff_geo}, with 
$\mathfrak{h} = \mathfrak{so}(1,3)$.  The first term in this 
expression is expanded as
\begin{displaymath}
	dA_\mathfrak{p} = d(i A^a P_a) 
	= i\, dA^a P_a - i \bigg(\frac{dl}{l} \wedge A^a \bigg) P_a,
\end{displaymath}
since $P_a = M_{a4}/l$. By use of the relation~\eqref{eq:rel.Ll} 
between $l$ and the cosmological function $\Lambda$, the last term 
of the torsion can be rewritten as
\begin{displaymath}
	- d\ln l \wedge A^a = \tfrac{1}{2} d\ln\Lambda \wedge A^a,
\end{displaymath}
which shows that this contribution depends on the relative 
infinitesimal change of the cosmological function along 
spacetime, rather than on its absolute change.

Although the curvature and torsion have contributions that are 
not there for a Riemann-Cartan geometry, the Bianchi identities 
are unchanged:
\begin{subequations}
	\label{eqs:lin_Bianchi}
	\begin{align}
	\label{eq:lin_Bianchi1}
	d_A \circ d_A A^{ab} 
	&\equiv 0,
	\\
	\label{eq:lin_Bianchi2}
	d_A \circ d_A A^a + A^b \wedge d_A A\ind{_b^a} 
	&\equiv 0,
	\end{align}
\end{subequations}
where $d_A$ is the exterior covariant derivative with respect to 
the spin connection.

The transformations that are consistent with the given geometry 
are local Lorentz transformations and spacetime diffeomorphisms, 
the latter being unphysical as they merely relabel spacetime 
coordinates~\cite{Edelstein:2006a}. In contrast, with respect to 
elements of $SO(1,4)$, we see from~\eqref{eq:red_split_AF_dSC} 
that the spin connection and vierbein, and the torsion and 
curvature form irreducible multiplets. Due to the reductive 
nature of $\mathfrak{so}(1,4)$, these geometric objects are well 
defined up to local Lorentz transformations only. Since local 
translational symmetry may play an important role in theories of 
gravity, there is the need to extend the structure group to 
$SO(1,4)$, while preserving the presence of these different 
objects, necessary to construct geometric theories of gravity.  
This will be discussed for the given de~Sitter-Cartan geometry in 
the following section.

\section{$\textbf{SO(1,4)}$-invariant de~Sitter-Cartan geometry 
	with a cosmological function}
\label{sec:nonlin_dSC_geo}

In order to extend the structure group to $SO(1,4)$ and have 
geometric objects that are well defined through the decomposition 
of a Cartan connection and curvature according to the reductive 
splitting~\eqref{eq:red_split_so(1,4)}, we nonlinearly realize the 
de~Sitter-Cartan connection of Sec~\ref{sec:lin_dSC_geo}. To 
realize connections on spacetime in a nonlinear way was first 
considered by Stelle and West~\cite{Stelle:1979va, 
	stelle.west:1980ds}, while its usefulness for theories of 
gravity has been pointed out in, e.g., \cite{Wise:2011-sym.br, 
	Tiemblo:2005js, Tresguerres:2008jf, Hehl:2013gtg}.  The 
formalism of nonlinear realizations was developed to 
systematically study spontaneous symmetry breaking in 
phenomenological field 
theory~\cite{Coleman:1969sm,Callan:1969sn,Volkov:1973vd}, in 
which linearly transforming irreducible multiplets become 
nonlinear but reducible realizations, when the symmetry group is 
broken to one of its subgroups.

A Cartan connection on a principal Lorentz bundle $P$ may be 
thought of as an Ehresmann connection on a principal 
$SO(1,4)$-bundle $Q$ over $\mathcal{M}$ that is reduced to 
$P$~\cite{sharpe1997diff_geo}. This is in essence a symmetry 
breaking process~\cite{Wise:2011-sym.br}, for the reason that it 
corresponds to singling out a section $\xi$ of the associated 
bundle $Q \times_{SO(1,4)} dS$ of tangent de Sitter spaces, 
thereby reducing the structure group $SO(1,4)$ pointwise to 
$SO(1,3)_\xi$, the isotropy group of the point $\xi(x)$ in the 
internal de~Sitter space 
$dS_x$~\cite{husemoller:1966fibre,gibbons:2009b}. Most 
importantly, the reduction is not canonical, i.e., the section 
$\xi$ can be chosen arbitrarily, and the broken symmetries are 
nonmanifestly restored by realizing them nonlinearly through 
elements of the Lorentz group. Consequently, decomposing a 
nonlinear de~Sitter-Cartan connection according to the reductive 
splitting of $\mathfrak{so}(1,4)$ gives way to true geometric 
objects, well defined with respect to all elements of $SO(1,4)$.

Before we construct a nonlinear de~Sitter-Cartan geometry with a 
nonconstant cosmological function, we recall a handful of facts on 
nonlinear realizations for the de~Sitter group, see 
also~\cite{Zumino1977189,wess:1992ss}. Within some neighborhood of 
the identity, an element $g$ of $SO(1,4)$ can uniquely be 
represented in the form
\begin{displaymath}
	g = \exp(i\xi\cdot P) \tilde{h},
\end{displaymath}
with $\tilde{h} \in SO(1,3)$ and $\xi\cdot P = \xi^a P_a$.
The $\xi^a$ parametrize the coset space $SO(1,4)/SO(1,3)$ so that 
they constitute a coordinate system for de~Sitter space. This 
parametrization allows us to define the action of $SO(1,4) \ni 
g_0$ on de~Sitter space by
\begin{displaymath}
	g_0 \exp(i\xi\cdot P) = \exp(i\xi'\cdot P)h'~;
	\quad
	h' = \tilde{h}'\tilde{h}^{-1},
\end{displaymath}
where $\xi'=\xi'(g_0,\xi)$ and $h'=h'(g_0,\xi)$ are in general 
nonlinear functions of the indicated variables. In case $g_0 = 
h_0$ is an element of $SO(1,3)$, the action is linear and the 
transformation of $\xi$ is given explicitly by
\begin{displaymath}
	h_{0}: i\xi\cdot P \mapsto i\xi'\cdot P = i\xi\cdot 
	\mathrm{Ad}(h_0)(P).
\end{displaymath}
If on the other hand $g_0 = 1 + i\epsilon\cdot P$ is an 
infinitesimal pure de~Sitter translation, the variations 
$\delta\xi^a$ and $\delta h^{ab}$ satisfy
\begin{multline*}
	\exp(-i\xi\cdot P) i\epsilon\cdot P \exp(i\xi\cdot P) \\
	- \exp(-i\xi\cdot P) \delta\!\exp(i\xi\cdot P) = \tfrac{i}{2} 
	\delta h \cdot M,
\end{multline*}
where $\tfrac{i}{2} \delta h\cdot M = h'-1 \in 
\mathfrak{so}(1,3)$ and $\delta h\cdot M = \delta h^{ab}M_{ab}$.
This equation is solved by
\begin{align}
	\label{eq:inf_tr_xi}
	\delta\xi^a &= \epsilon^a + \Big(\frac{z\cosh z}{\sinh z} - 
	1\Big) \bigg(\epsilon^a - \frac{\xi^a \epsilon_b 
		\xi^b}{\xi^2}\bigg), \\
	\label{eq:inf_tr_h}
	\delta h^{ab} &= \frac{1}{l^2} \frac{\cosh z - 1}{z\sinh z} 
	(\epsilon^a\xi^b - \epsilon^b\xi^a),
\end{align}
where we made use of the notation $z = l^{-1} \xi$ and $\xi = 
(\eta_{ab} \xi^a \xi^b)^{1/2}$.

Subsequently, let $\psi$ be a field that belongs to some linear 
representation $\sigma$ of $SO(1,4)$. Given a local section of 
the associated bundle of homogeneous de~Sitter spaces, i.e.,~$\xi: 
U \subset \mathcal{M} \to U \times dS$, the corresponding 
nonlinear field is constructed pointwise as 
\begin{equation}
	\label{eq:def_nonlinear_field}
	\bar{\psi}(x) = \sigma(\exp(-i\xi(x)\cdot P))\psi(x).
\end{equation}
Under a local de~Sitter transformation $g_0$, it rotates only 
according to its $SO(1,3)$-indices, namely, $\bar{\psi}'(x) = 
\sigma(h'(\xi,g_0)) \bar{\psi}(x)$. It is manifest that the 
irreducible linear representation~$\psi$ has given way to a 
nonlinear and reducible realization~$\bar{\psi}$.

In concordance with the 
prescription~\eqref{eq:def_nonlinear_field} to construct 
nonlinear realizations, the nonlinear $\mathfrak{so}(1,4)$-valued 
Cartan connection is defined as~\cite{stelle.west:1980ds}
\begin{equation}
	\label{eq:A_nonlin}
	\bar{A} = \mathrm{Ad}(\exp(-i\xi\cdot P))(A + d).
\end{equation}
Under local de~Sitter transformations, the field $\bar{A}$ 
transforms according to
\begin{displaymath}
	\bar{A} \mapsto \Ad(h'(\xi,g_0))(\bar{A}+d).
\end{displaymath}
Because elements of $SO(1,4)$ are nonlinearly realized as 
elements of $SO(1,3)$, the reductive decomposition 
$\bar{A}_\mathfrak{h} + \bar{A}_\mathfrak{p}$ is invariant under 
local de~Sitter transformations.  It is then sensible to define 
the spin connection and vierbein through these projections as 
$\omega = \bar{A}_\mathfrak{h}$ and $e = \bar{A}_\mathfrak{p}$, 
respectively.

The spin connection $\omega$ and vierbein $e$ can be expressed in 
terms of the section $\xi$ and the projections $A_\mathfrak{h}$ 
and $A_\mathfrak{p}$ of the linear $SO(1,4)$ connection.  These 
relations follow from~\eqref{eq:A_nonlin}, in which the different 
objects appear according to
\begin{multline*}
	\tfrac{i}{2} \omega^{ab} M_{ab} + i e^a P_a \\
	= \Ad(\exp(-i\xi\cdot P)) \Big( \tfrac{i}{2} A^{ab} M_{ab} + i 
	A^a P_a + d \Big).
\end{multline*}
To carry out the computation of the right-hand side we utilize the 
techniques of~\cite{Zumino1977189,stelle.west:1980ds}, explained 
in their appendices. In short, one expands the adjoint action of 
the exponential as a power series in the adjoint action of its 
generating element $-i\xi\cdot P$. The latter is just the Lie 
commutator and is given explicitly
in~\eqref{eq:comm_relations_so(1,4)}. We find
\begin{widetext}
\begin{subequations}
\label{eqs:nonlin_spin_vier}
\begin{align}
\label{eq:nonlin_spinconn}
	\omega^{ab} &= A^{ab} - \frac{\cosh z - 1}{l^2 z^2} \big[ 
	\xi^a (d\xi^b + A\ind{^b_c} \xi^c) - \xi^b (d\xi^a + 
	A\ind{^a_c} \xi^c) \big] - \frac{\sinh z}{l^2 z} (\xi^a A^b - 
	\xi^b A^a),
\\
\label{eq:nonlin_vierbein}
	e^a &= A^a + \frac{\sinh z}{z}( d\xi^a + A\ind{^a_b} \xi^b) - 
	\frac{dl}{l} \xi ^a 
	+ (\cosh z - 1) \bigg( A^a - \frac{\xi^b A_b \xi^a}{\xi^2} 
	\bigg) - \bigg( \frac{\sinh z}{z} - 1 \bigg) \frac{\xi^b 
		d\xi_b \xi^a}{\xi^2}.
\end{align}
\end{subequations}
\end{widetext}
These expressions are almost identical to the corresponding 
objects found by Stelle and West~\cite{stelle.west:1980ds}. The 
difference to note is that we have a new term in the 
expression~\eqref{eq:nonlin_vierbein} for the vierbein, namely, 
$-l^{-1} dl\,\xi^a$. This term is present because it is possible 
that the internal de Sitter spaces are characterized by 
cosmological constants that are not necessarily equal along 
spacetime. More precisely, one has to take into account the 
possibility that the in $\mathfrak{p}$ defined length scale is a 
nonconstant function, see~Sec.~\ref{sec:lin_dSC_geo}.  On the 
other hand, the results of~\cite{stelle.west:1980ds} specialize 
for the case that the local de Sitter spaces have the same 
pseudo-radius at any point in spacetime. When $l$ is a constant 
function, one naturally recovers the results 
of~\cite{stelle.west:1980ds}. 

Upon the action of local de Sitter transformations, the linear 
curvature $F$ rotates in the adjoint representation. Therefore, 
one deduces that the nonlinear Cartan curvature $\bar{F}$ is 
equal to the exterior covariant derivative of the nonlinear 
connection, i.e.,
\begin{equation}
\label{eq:F_nonlin}
	\bar{F} = \Ad(\exp(-i\xi\cdot P))(F) = d\bar{A} + 
	\tfrac{1}{2}[\bar{A},\bar{A}],
\end{equation}
which complies with the structure of a Cartan geometry.
The nonlinear Cartan curvature is a $\mathfrak{so}(1,4)$-valued 
$2$-form on spacetime, which we decompose once again according 
to~$\bar{F} = \bar{F}_\mathfrak{h} + \bar{F}_\mathfrak{p}$. Since 
$\bar{F}$ transforms---in general, nonlinearly---with elements of 
$SO(1,3)$, the reductive splitting is invariant under local 
de~Sitter transformations. This suggests that 
$\bar{F}_\mathfrak{h}$ and $\bar{F}_\mathfrak{p}$ must be 
considered the genuine curvature and torsion of the Cartan 
geometry, which are denoted by $R$, respectively $T$. The 
definition~\eqref{eq:F_nonlin} implies that
\begin{multline*}
	\tfrac{i}{2} R^{ab} M_{ab} + i T^a P_a \\
	= \Ad(\exp(-i\xi\cdot P)) \Big( \tfrac{i}{2} F^{ab} M_{ab} + i 
	F^a P_a \Big),
\end{multline*}
from which one is able to express the curvature and torsion in 
terms of $\xi$, $F_\mathfrak{h}$ and 
$F_\mathfrak{p}$:
\begin{subequations}
\label{eqs:nonlin_curv_tors}
\begin{align}
	\nonumber
	R^{ab} &= F^{ab} - \frac{\cosh z - 1}{l^2 z^2}\, \xi^c (\xi^a 
	F\ind{^b_c} -  \xi^b F\ind{^a_c}) \\
	\label{eq:nonlin_curv}
	&\hspace{3cm} - \frac{\sinh z}{l^2 z} (\xi^a F^b - \xi^b F^a),
\\
\label{eq:nonlin_tors}
	T^a &= \frac{\sinh z}{z} \xi^b F\ind{^a_b} + \cosh z\, F^a + 
	(1 - \cosh z) \frac{\xi_b F^b \xi^a}{\xi^2}.
\end{align}
\end{subequations}
From~\eqref{eq:F_nonlin} it furthermore follows that
\begin{displaymath}
	R^{ab}
	= d_\omega \omega^{ab} + \frac{1}{l^2} e^a \wedge e^b
	\quad\text{and}\quad
	T^a
	= d_\omega e^a - \frac{1}{l} dl \wedge e^a.
\end{displaymath}
These equations, which express the curvature and torsion in terms 
of the spin connection and vierbein, are the ones expected for a 
Cartan geometry.
Because the exterior covariant derivative of $\bar{F}$ is always 
zero, there are two Bianchi identities that are formally the same 
as those given by~\eqref{eqs:lin_Bianchi},~i.e.,
\begin{displaymath}
	d_\omega \circ d_\omega \omega^{ab} \equiv 0
	\quad\text{and}\quad
	d_\omega \circ d_\omega e^a + e^b \wedge d_\omega 
	\omega\ind{_b^a} \equiv 0.
\end{displaymath}

When the section $\xi$ is gauge-fixed along spacetime, and for 
convenience at any point be chosen the origin of the tangent 
de~Sitter spaces, i.e.,~$\xi^a(x) = 0$, all the expressions reduce 
to those of Sec.~\ref{sec:lin_dSC_geo}.  This is to be expected, 
because the broken symmetries are not considered, and the geometry 
is described simply by a $SO(1,4)$ Ehresmann connection for which 
only $SO(1,3)$-transformations---the isotropy group of $\xi^a 
=0$---are taken into account. This has precisely been the way in 
which the de~Sitter-Cartan geometry of Sec.~\ref{sec:lin_dSC_geo} 
was set up.

Finally, let us remark that if $A^{ab}$ and $A^a$ can be made to 
vanish everywhere, so that also $F^{ab}$ and $F^a$ are equal to 
zero, it follows that
\begin{displaymath}
	R^{ab} = 0 \quad\text{and}\quad T^a = 0.
\end{displaymath}
This shows that the nonhomogeneity of $\mathcal{M}$ is encoded in 
$A$ and $F$, and naturally independent of the section $\xi$.

\section{Conclusions and outlook}
\label{sec:conclusion}

In this work we have generalized the geometric framework of 
de~Sitter-Cartan spacetimes with a cosmological constant to the 
case of a nonconstant cosmological function $\Lambda$. A 
de~Sitter-Cartan spacetime consists of a principal Lorentz bundle 
over spacetime, on which is defined a $\mathfrak{so}(1,4)$-valued 
Cartan connection. It accounts for a spin connection and 
vierbein, as well as for their curvature and torsion, whereas 
spacetime is locally approximated by de~Sitter spaces. The 
cosmological constants of these tangent de~Sitter spaces are 
determined by a length scale, defined in the translational part 
of $\mathfrak{so}(1,4)$. By letting this length scale depend 
arbitrarily on the spacetime point in Sec.~\ref{sec:lin_dSC_geo}, 
we obtained a de~Sitter-Cartan geometry that accommodates a 
cosmological function by construction. Most importantly, it was 
shown that a nonconstant $\Lambda$ gives rise to an extra 
contribution in the expression for the torsion, in which the 
cosmological function appears through its logarithmic derivative.  
In the limit $\Lambda \to 0$ one recovers the well-known 
Riemann-Cartan spacetime with arbitrary curvature and torsion.
In Sec.~\ref{sec:nonlin_dSC_geo}, the de~Sitter-Cartan connection 
has been realized nonlinearly in order to obtain 
$SO(1,4)$-covariant definitions for the spin connection and 
vierbein, and likewise for the curvature and torsion. This 
generalized previous results to include a nonconstant $\Lambda$.

The cosmological function could be used to model dark energy that 
changes along space and time, in which way it might give an 
alternative description for one of the models for time-evolving 
dark energy~\cite{Peebles:2003cc,Copeland:2006de}. To determine 
the value of the cosmological function along spacetime, an 
adequate action for the gravitational field coupled to matter 
will have to be defined. By including invariants of the torsion 
tensor on the gravitational side, the first derivative of the 
cosmological function will automatically be present. Matter 
fields can be coupled both minimally to the gravitational field 
or in another nonminimal way. The analysis of such models that 
make use of the framework outlined in this paper is an 
interesting and important subject of future research. Since the 
cosmological function quantifies the lack of commutation of two 
infinitesimal spacetime translations, the local kinematics on 
spacetime depend on $\Lambda$. Consequently, there would be a 
link between the dynamical character of the cosmological function 
and its kinematical implications.

Another point of interest comes about upon noting that, when the 
de~Sitter algebra is contracted to the Poincar\'e algebra, 
namely, when $l \to \infty$ in the commutation 
relations~\eqref{eq:comm_relations_so(1,4)}, the geometric 
objects of Sec.~\ref{sec:nonlin_dSC_geo} reduce to those of 
teleparallel gravity~\cite{aldrovandi:2012tele,Arcos:2005ec}.  
This observation suggests that the geometry of spacetime that 
underlies teleparallel gravity is described by a Riemann-Cartan 
geometry (with vanishing curvature), for which the Poincar\'e 
translations are realized nonlinearly as elements of $SO(1,3)$.  
In fact, from~\eqref{eq:inf_tr_h} one sees that the nonlinear 
element of the Lorentz algebra, which corresponds to an 
infinitesimal Poincar\'e translation with parameters 
$\epsilon^a$, vanishes, for
\begin{displaymath}
	\delta h^{ab} = \lim_{l \to \infty} \frac{1}{l^2} \frac{\cosh 
		z - 1}{z\sinh z} (\epsilon^a\xi^b - \epsilon^b\xi^a) = 0.
\end{displaymath}
One then concludes that any Poincar\'e translation is trivially 
realized by the identity transformation, a property that is 
relied upon in the interpretation of teleparallel gravity as a 
gauge theory for the Poincar\'e translations. Given the knowledge 
that the geometric structure of teleparallel gravity is such a 
Riemann-Cartan spacetime, the de~Sitter-Cartan geometry of 
Sec.~\ref{sec:nonlin_dSC_geo} might be the right framework to 
generalize teleparallel gravity to a theory that is invariant 
under local $SO(1,4)$-transformations, in place of the elements 
of the Poincar\'e group.

\begin{acknowledgments}

The author would like to thank J.~G.~Pereira for helpful 
discussions and suggestions. He also gratefully acknowledges 
financial support by CAPES.

\end{acknowledgments}


%

\end{document}